\documentclass[prl,twocolumn,floatfix,amsfonts]{revtex4}
\usepackage{graphicx,graphics,color,epsfig}
\usepackage{bm}
\usepackage{amsfonts}
\usepackage{amsmath}
\usepackage{amssymb}
\usepackage{epstopdf}

\newcommand{\beq}{\begin{equation}}
\newcommand{\eeq}{\end{equation}}
\newcommand{\beqa}{\begin{eqnarray}}
\newcommand{\eeqa}{\end{eqnarray}}

\begin{document}

\title{A non-linear transport method for detecting superconducting stripes}

\author{Rodrigo A. Muniz$^{1,2}$ and Ivar Martin$^2$}
\affiliation{$^1$Department of Physics and Astronomy, University of Southern California, Los Angeles, CA 90089\\
$^2$Theoretical Division, Los Alamos National Laboratory, Los Alamos, New Mexico 87545}

\begin{abstract}
We theoretically study the effect of stripe-like superconducting inclusions on the non-linear resistivity in single crystals. Even when the stripe orientation varies throughout the sample between two orthogonal directions due to twinning, we predict that there should be a universal scaling relationship between the nonlinear resistivity curves measured at different angles relative to the crystal axes. This prediction can be used to verify or rule out the existence of superconducting stripes at and above the superconducting transition temperature in cuprate superconductors.
\end{abstract}

\maketitle

Study of the high-temperature superconductivity (HTS) problem has been complicated by proximity of the superconducting state to non-superconducting ordered phases. Among them, only antiferromagnetic insulating state has been unambiguously identified as always present in the undoped parent compounds across all families of cuprates. The nature of the pseudogap regime \cite{Timusk} that emerges upon doping, has remained elusive, even though there are many theoretical proposals for its origin \cite{RVB, Zaanen, Kivelson, DDW, Varma}. Among the simplest, from the stand point of detection, is the possibility that the pseudogap regime contains a region of spin and/or charge density waves (SDW/CDW), or their strongly-coupled cousin, the stripe state \cite{Zaanen,Kivelson}. Indeed, neutron scattering commonly detects elastic and inelastic responses characteristic of static or slowly fluctuating incommensurate SDW in majority of cuprate families \cite{Tranquada, Hinkov}. Charge modulation is also observed \cite{STM, Huecker, Yazd}, albeit less commonly \cite{Abba}. 
Naturally, charge and spin modulation can locally modify the conditions for  onset of superconductivity. Moreover, it is possible that in the anharmonic regime that characterizes the stripe state, local pairing mechanisms, which are not present in the more common CDW and SDW setting, may become operational \cite{EKZ, ACN, MPK}, in which case stripes would be an important contributor to the high values of superconducting transition temperature $T_c$. Inhomogeneous superconductivity could naturally account for suppressed  superfluid density \cite{Uemura} and for spectral weight transfer observed in the optical conductivity of underdoped cuprates \cite{Oren2}.

Identification of the stripe states, in contrast to harmonic spin or charge modulations has been notoriously difficult. Direct experiments that would measure the transport anisotropy induced by the rotational symmetry breaking caused by stripes are often hindered by sample twinning or stripe ordering patterns that restore rotational symmetry on the measurement length scale \cite{Ando}. 
In addition, it is believed that in order to be compatible with superconductivity, the stripes have to be dynamical, i.e., fluctuating on some timescale. 

In this letter we propose a method capable of detecting signatures of  superconducting stripes diluted in a normal matrix despite all these complications.
It is based on probing the spatial anisotropy of the {\em non-linear transport}, which should be induced by superconducting stripes in single crystal samples, even when linear transport is completely isotropic.  The method takes advantage of the fact that unlike the linear response conductivity tensor that has to be rotationally invariant in tetragonal systems, the non-linear response in general is not. Superconductors have inherently non-linear $I$-$V$ characteristics, which makes them ideally suited for the proposed method.  Therefore, non-linear transport measurements can help both to establish the temperature range within which superconducting inclusions persist above above $T_c$, as well as to determine the local spatial structure of these inclusions.

\begin{figure}[ht]
\begin{tabular}{ll}
{\includegraphics[width=1.6in]{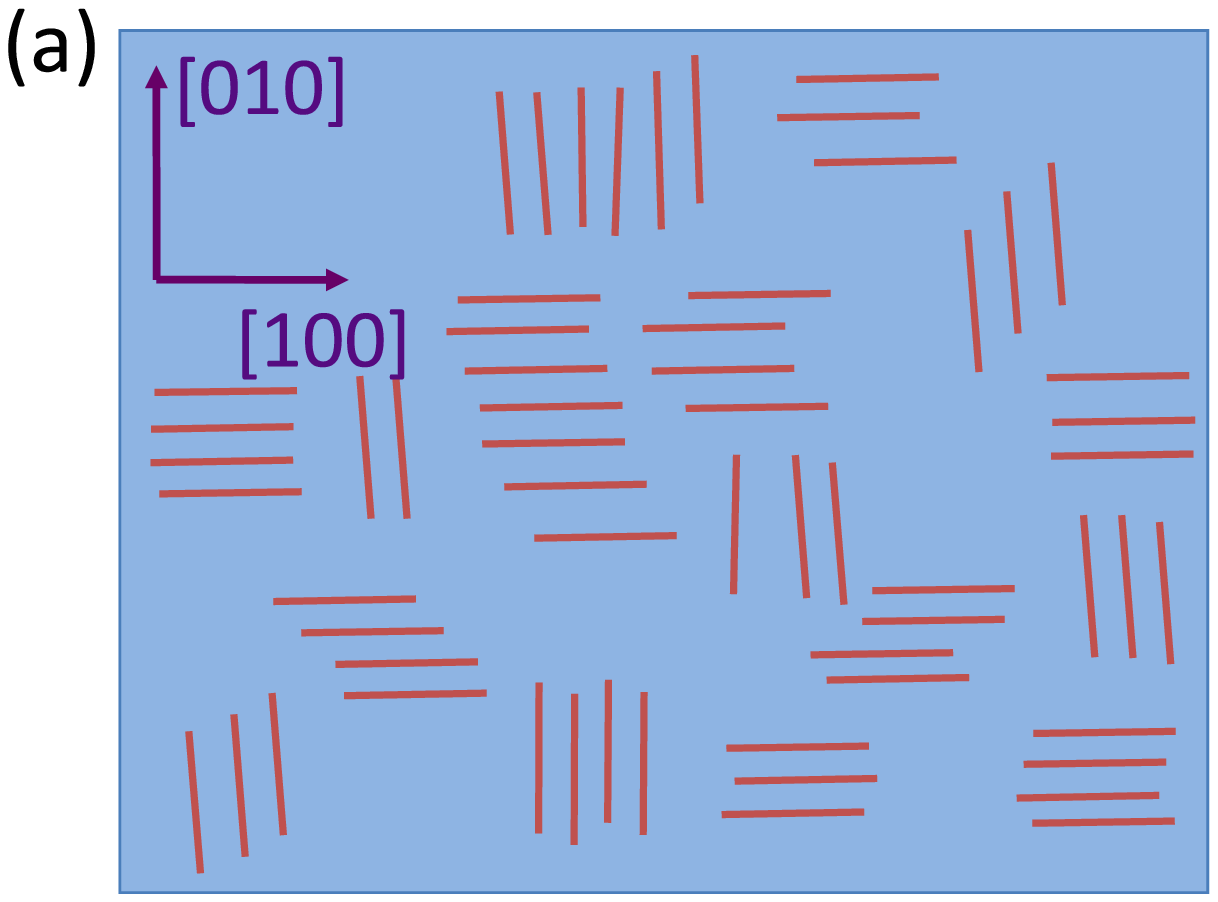}}  & {\includegraphics[width=1.8in]{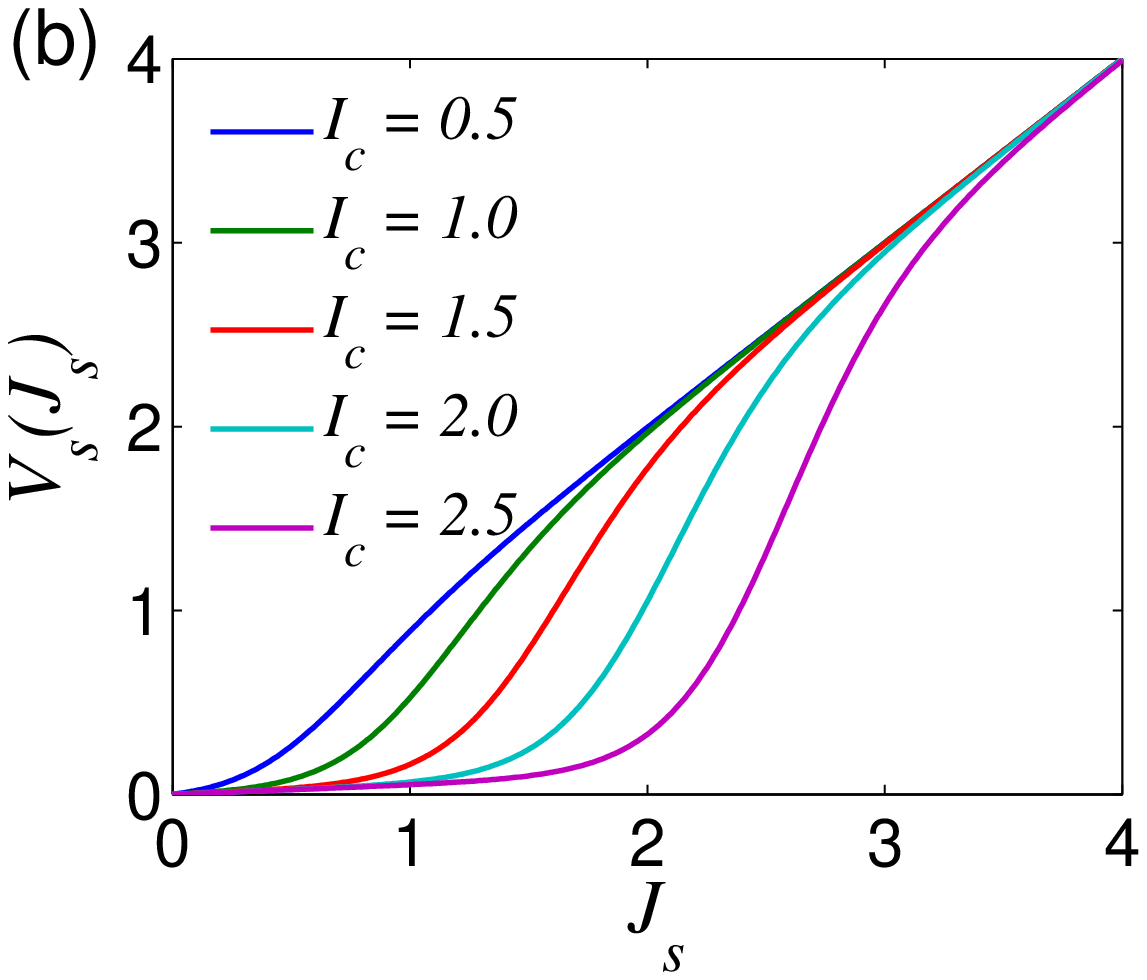}} \\
\end{tabular}
\caption{(a) System with stripes domains aligned along directions [100] and [010]. (b) $I$-$V$ relations used for the superconducting stripes for different values of the critical current  $I_c $.}
\label{fig:system} 
\end{figure}

We qualitatively demonstrate how anisotropic non-linear resistivity can arise in a tetragonal system by considering a schematic model in Fig.~\ref{fig:system}a.
Suppose that within the CuO$_2$ planes there are domains of superconducting stripes, which are primarily oriented along $[100]$ and $[010]$ directions. 
Each superconducting segment has a non-linear $I$-$V$ characteristic, as shown schematically in Fig.~\ref{fig:system}b.
If the characteristic domain size is much smaller than the crystal size, the tetragonal symmetry of the system is not broken -- the [100] direction is still equivalent to [010]. Then, we are immediately forced to conclude that linear resistivity must remain isotropic. In particular $\rho^0_{[110]} = \rho^0_{[100]}$ (superscript ``0'' indicates that the resistivity is taken at zero current).

Now, let us consider the non-linear resistivity of the system. Suppose we apply current $j_x$ and measure electric field  $E_x$, with direction $\hat x$ either parallel to [100] or [110] (two high-symmetry directions). For small applied current, the superconducting segments have negligible resistivity, and hence average resistivity of the system the smallest. As $j_x$ increases, some of the segments quench (when the local critical current $I_c$ is reached), and the overall resistivity increases. We therefore anticipate that in both cases, the resistivity will increase non-linearly with $j_x$.
However, the nonlinear resistivity in two cases will not be the same, as can be easily seen from the following argument, which is particularly simple in the dilute stripe concentration limit.
Suppose that in addition to the current in the direction $\hat x$, there is also a current in the $\hat y$-direction. Qualitatively, it is clear that it should make little difference for $E_x$ in case $\hat x = [100]$, but a big difference when $\hat x = [110]$.  That is because, in the latter case, $j_y$ current will directly contribute to reaching the critical current $I_c$ on the SC links involved in the transport along $\hat x$, while in the former, $j_y$ primarily saturates the horizontal superconducting links, which are largely irrelevant for the transport along $\hat x$.
Thus we see that the non-linear resistivity has to be anisotropic, and that the $I$-$V$ dependence in the coordinate system where $\hat x = [100]$ and $\hat y = [010]$ with good accuracy is given by 
\beqa
E_x = \rho(j_x) j_x,\nonumber\\
E_y = \rho(j_y) j_y.\label{eq:Ej}
\eeqa
That is, the electric field along the stripe pinning direction depends only on the current in the same direction.

In order to obtain the relation between the $\bf E$ and $\bf j$ in any other planar coordinate system we only need to perform a rotation of Eqs. (\ref{eq:Ej}) around $\hat z$-axis by angle $\theta$ with the matrix ${\cal R}_{\theta}^{\hat{z}}$,
\begin{equation}
{\bf E'}={\cal R}_{\theta}^{\hat{z}}{\bf \rho}({\bf J}){\bf J}={\cal R}_{\theta}^{\hat{z}}{\bf \rho}({\cal R}_{-\theta}^{\hat{z}}{\bf J^{\prime}}){\cal R}_{-\theta}^{\hat{z}}{\bf J^{\prime}}
\end{equation}
where
\begin{equation}
{\bf \rho}({\bf J}) \equiv 
\left[\begin{array}{cc}
\rho(J_{x}) & 0\\
0 & \rho(J_{y})
\end{array}\right].
\end{equation}
The resistivity tensor in the rotated basis is 
\begin{equation}
{\rho'}({\bf J^{\prime}})={\cal R}_{\theta}^{\hat{z}}{\bf \rho}({\cal R}_{-\theta}^{\hat{z}}{\bf J^{\prime}}){\cal R}_{-\theta}^{\hat{z}}.
\end{equation}
Naturally, in the limit of vanishing current we see that  the resistivity tensors coincide, $\rho' = \rho$, as expected from linear response of a tetragonal system. However, at a finite current, the relationship is more complex. For instance, if a current $j'$ is applied along direction $\hat x'$, the electric field response becomes
\begin{equation}
\begin{array}{rl}
E_{x'}= & [\rho(j^{\prime}\cos\theta)\cos^{2}\theta+\rho(j^{\prime}\sin\theta)\sin^{2}\theta] \,j'\\
E_{y'} =&   \left[\rho(j^{\prime}\cos\theta)-\rho(j^{\prime}\sin\theta)\right]\sin\theta\cos\theta\, j'\end{array}\label{eq:EjTheta}
\end{equation}
In particular, for the high symmetry direction  $\theta={\pi}/{4}$ the resistivity is obtained by dilation of the current axis by a $\sqrt{2}$ factor, $\rho'(j^{\prime})=\rho({j^{\prime}}/{\sqrt{2}})$.  Another notable feature is the appearance of the transverse electric field if current is applied away from the high-symmetry directions, i.e. when $\theta$ is not an integer multiple of $\pi/4$.  The induction of transverse electric field in tetragonal systems is only possible in the presence of nonlinear response.

To test the above reasoning, we consider an explicit resistor network model that incorporates the described qualitative features. The resistors, connecting the nearest neighbor sites of a square lattice, $R_{ij}$, are chosen at random to be either ``normal'', with constant resistivity $R_n$, or ``superconducting'', with the current-dependent resistivity $R_s(J_s)$. The probability of superconducting bonds is $p$.
The Kirchhoff's equations
\begin{equation}
\sum_{j=n.n.}\frac{V_{i}-V_{j}}{R_{ij}(V_{i}-V_{j})}=0,
\label{eqn:linsys}
\end{equation}
combined with current or voltage boundary conditions  give a system of nonlinear equations that can be used to determine the local voltages $\{V_i\}$.
In our model calculation, for the superconducting resistors we take the $I$-$V$ dependence  as
\begin{equation}
V_{s} = R_{s}(J_{s})J_{s} =\left(R_{s}^{0} + \frac{R_{n}-R_{s}^{0}}{e^{-4(J_{s}-I_c)}+1}\right) J_{s}
\end{equation}
which at small currents has resistivity approximately $R_{s}^{0}< R_n$ and at large currents $J_s >> I_c$ saturates to the resistance of the normal links $R_n$ as illustrated in Fig.~\ref{fig:system}b. We will take $R_n = 1$ and $R_s^0 = 0.05$ unless stated otherwise.  In the simulations we used a system of size $80\times 80$ sites with fixed voltage boundary conditions in the direction of applied current, and periodic boundary conditions in the transverse direction.  Simulations were performed for current applied along the bond and diagonal directions. 

\begin{figure}[ht]
\begin{tabular}{ll}
{\includegraphics[width=1.7in]{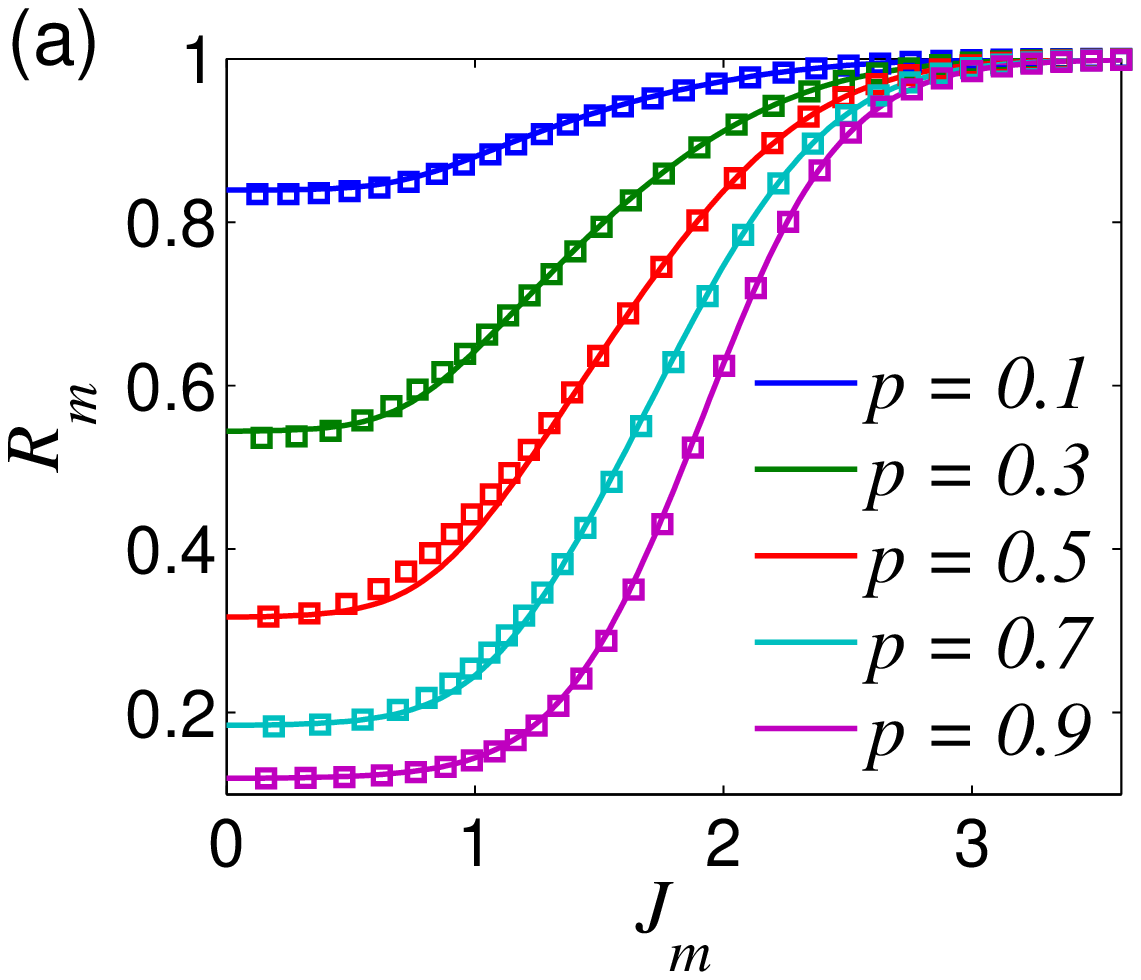}}  & {\includegraphics[width=1.7in]{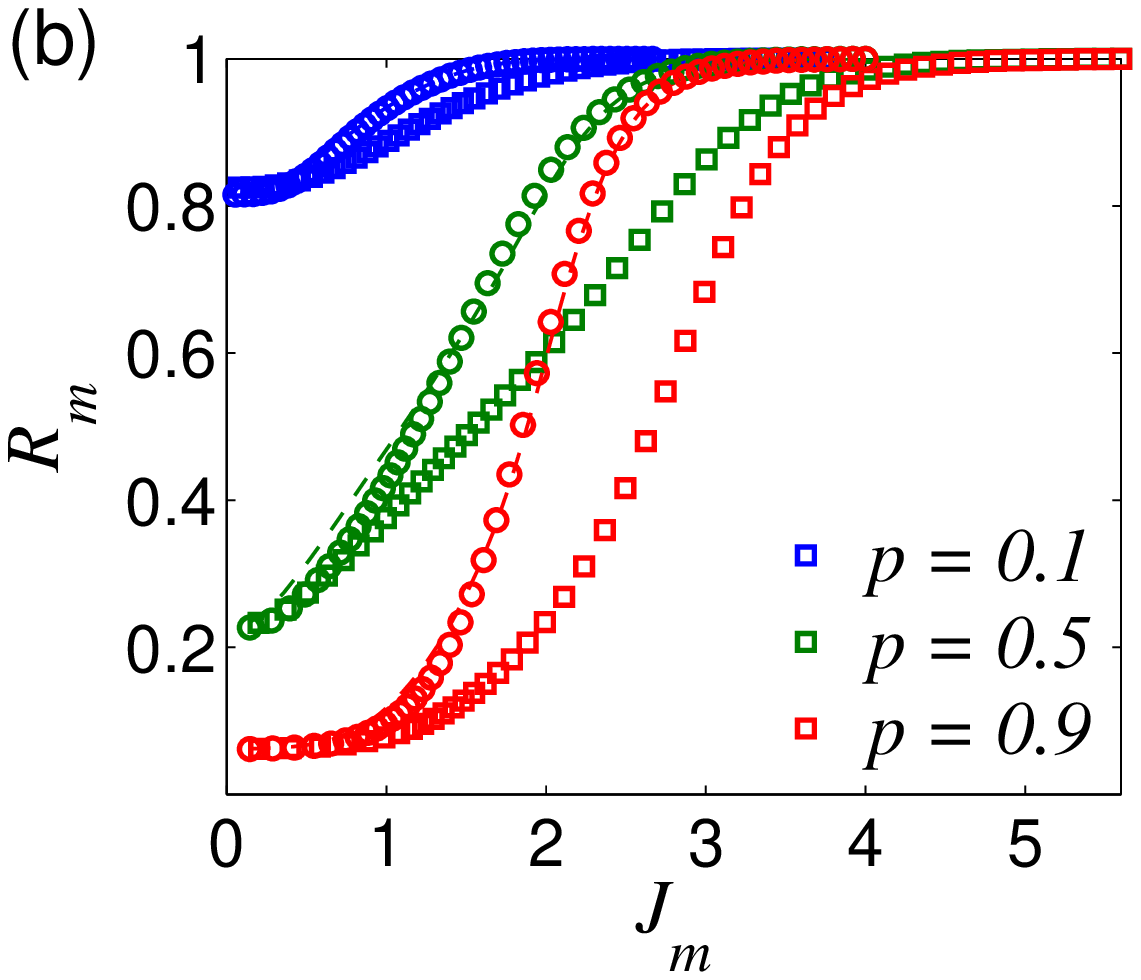}} 
\end{tabular}
\caption{(a) Comparison between EMT and the numerical lattice simulation. The lines correspond to Eq. (\ref{eq:EMT}) and the squares to the numerical simulation. (b) Verification of the ``$\sqrt{2}$ scaling''. 
Circles and squares correspond to numerical lattice simulations for the [100] and [110] directions, respectively. The dashed lines correspond to the numerical data for direction [110] with the current axis divided by $\sqrt 2$.}
\label{fig:compare} 
\end{figure}

In Fig.~\ref{fig:compare}a we present the result of lattice simulation (squares) when current is applied along the bond direction for various values of superconducting link concentrations $p$ and for $I_c = 2$. As expected, for large values of applied current $J_m$, the resistivity saturates to the ``normal" value 1, while at  low currents it drops by an amount that increases with $p$. The width of the intermediate nonlinear region decreases with increasing $p$. 

Our lattice simulation results can be very accurately reproduced by the effective medium theory (EMT).
EMT is typically applied to study {\em linear} random resistor networks, where it well captures, e.g., percolation phenomena \cite{Kirkpatrick}.
Within EMT, one solves exactly the problem of a particular random resistor $R_0$ embedded in a perfect matrix composed of identical effective medium resistors $R_m$. The value of $R_m$ is determined self-consistently from the condition that the voltage drop $V_0$ on resistor $R_0$, averaged over the distribution, be equal to the average voltage drop $V_m$ per link of the network. Here we apply this procedure to the binary network that contains nonlinear resistors. With the average current parallel to the bond direction, it leads to the following equations 
\beqa
J_s &=& \frac{2g_s(J_s)}{g_m(J_s) + g_s(J_s)}J_m \label{eq:J_s}\\
g_m(J_s) &=& (p - \frac{1}{2})(g_s(J_s) - g_n) \nonumber\\
&+& \left[(p - \frac{1}{2})^2( g_s(J_s)- g_n)^2 + g_n  g_s(J_s)\right]^{1/2} \label{eq:EMT}
\eeqa
where $J_m$ is the externally applied uniform current per unit cell,  $J_s$ is the current through a superconducting link, which needs to be determined self-consistently, and for convenience we introduced conductances, $g_i \equiv 1/R_i$.
These coupled nonlinear equations can be solved numerically for any specified form of $R_s(J)$.
As can be seen from Fig.~\ref{fig:compare}a, non-linear EMT very well approximates our lattice simulation results, with the largest deviation occurring at the percolation threshold, $p = 0.5$.

In Fig.~\ref{fig:compare}b we present the comparison of lattice simulations for the current applied along the bond (circles) and diagonal (squares) directions. After rescaling the current axis by a factor $\sqrt 2$, the diagonal resistivity (dashed lines) matches the resistivity in the bond direction, as was anticipated from Eq.~(\ref{eq:EjTheta}). Even at the percolation threshold, $p = 0.5$, the scaling works remarkably well. The deviation occurs due to the cross-talk between the horizontal and vertical superconducting bonds, which was neglected in Eq.~(\ref{eq:Ej}).
We have also performed simulations with ``longer" superconducting links with lengths of 2 to 4 lattice constants, which also confirmed $\sqrt 2$ scaling.

\begin{figure}[ht]
\begin{tabular}{ll}
{\includegraphics[width=1.7in]{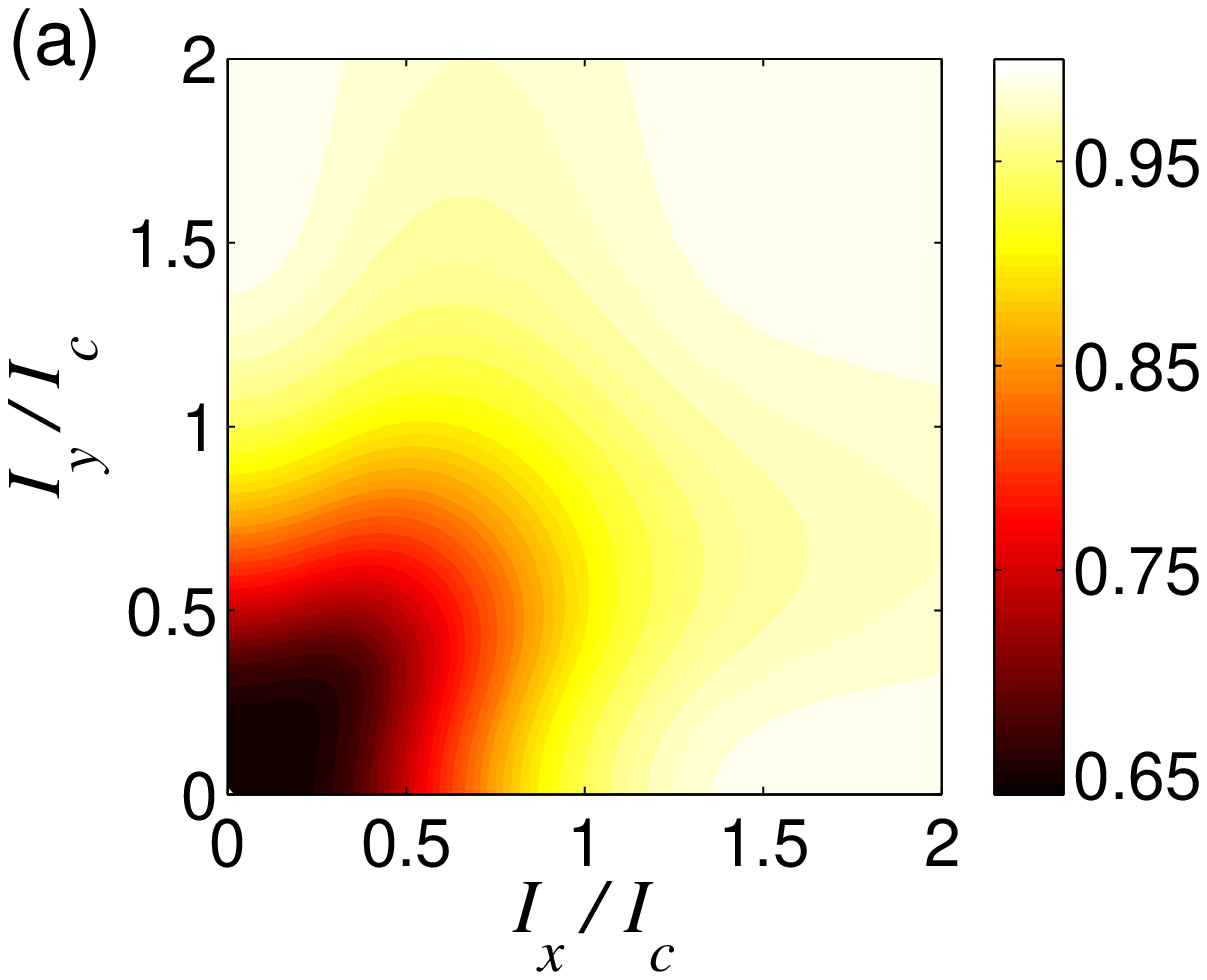}}  & {\includegraphics[width=1.7in]{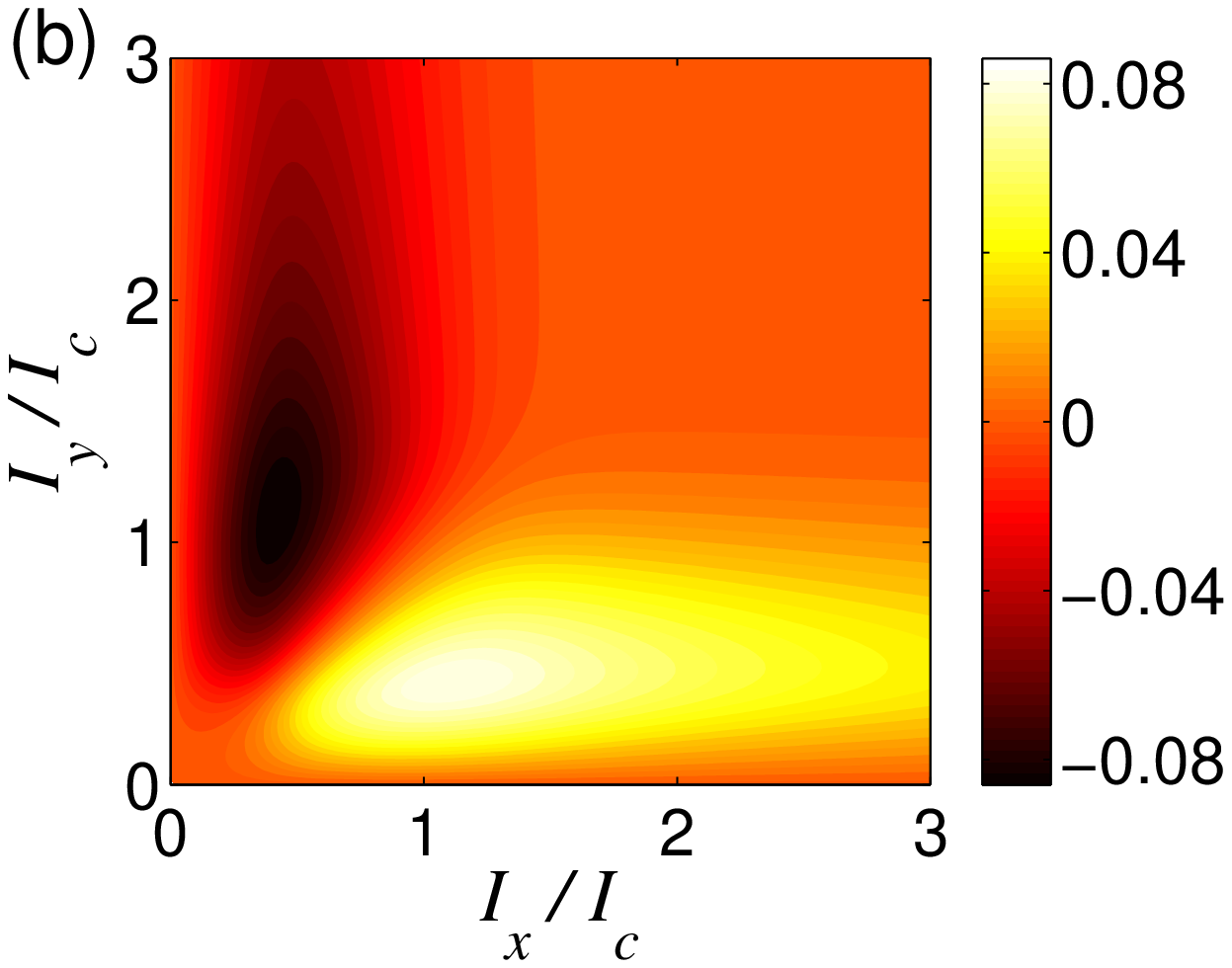}} \\
{\includegraphics[width=1.7in]{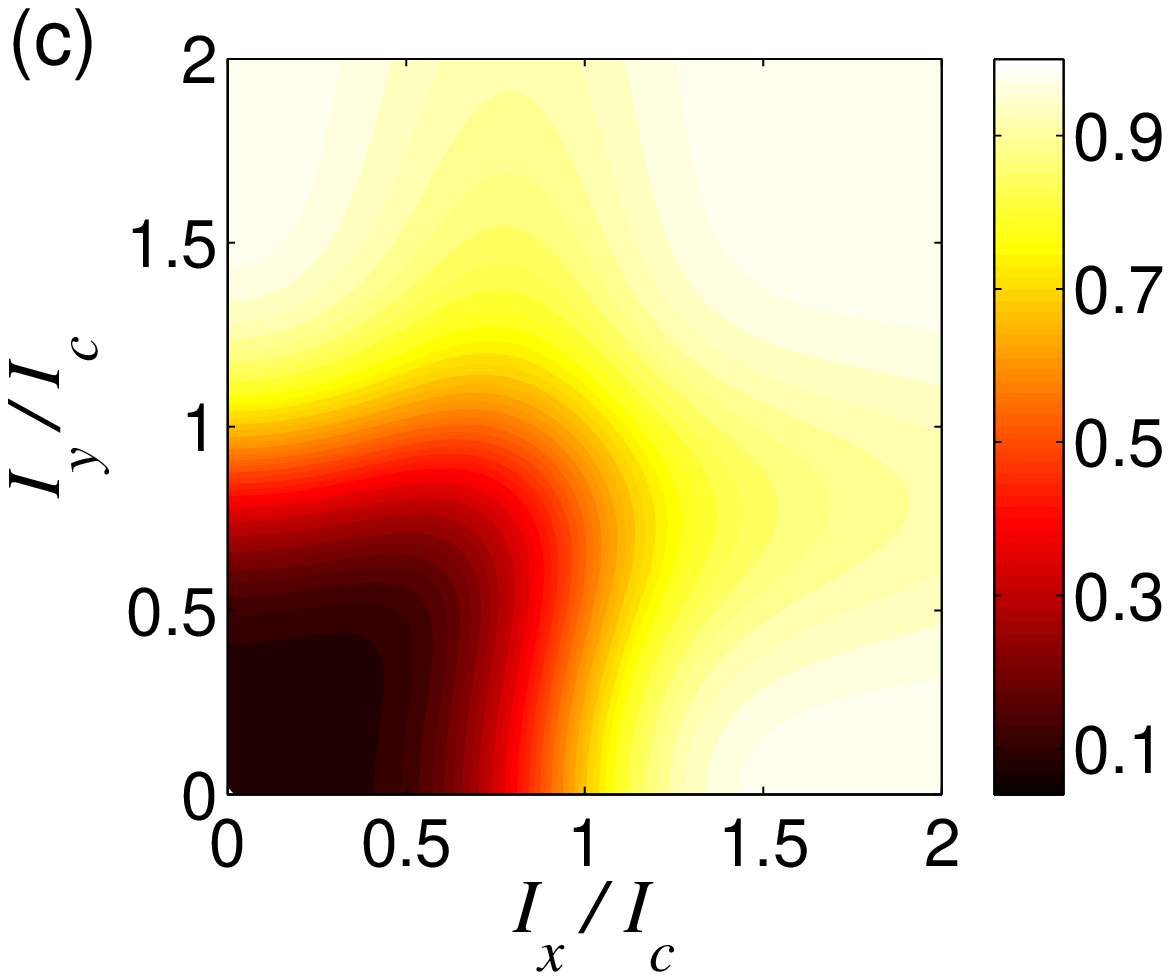}}  & {\includegraphics[width=1.7in]{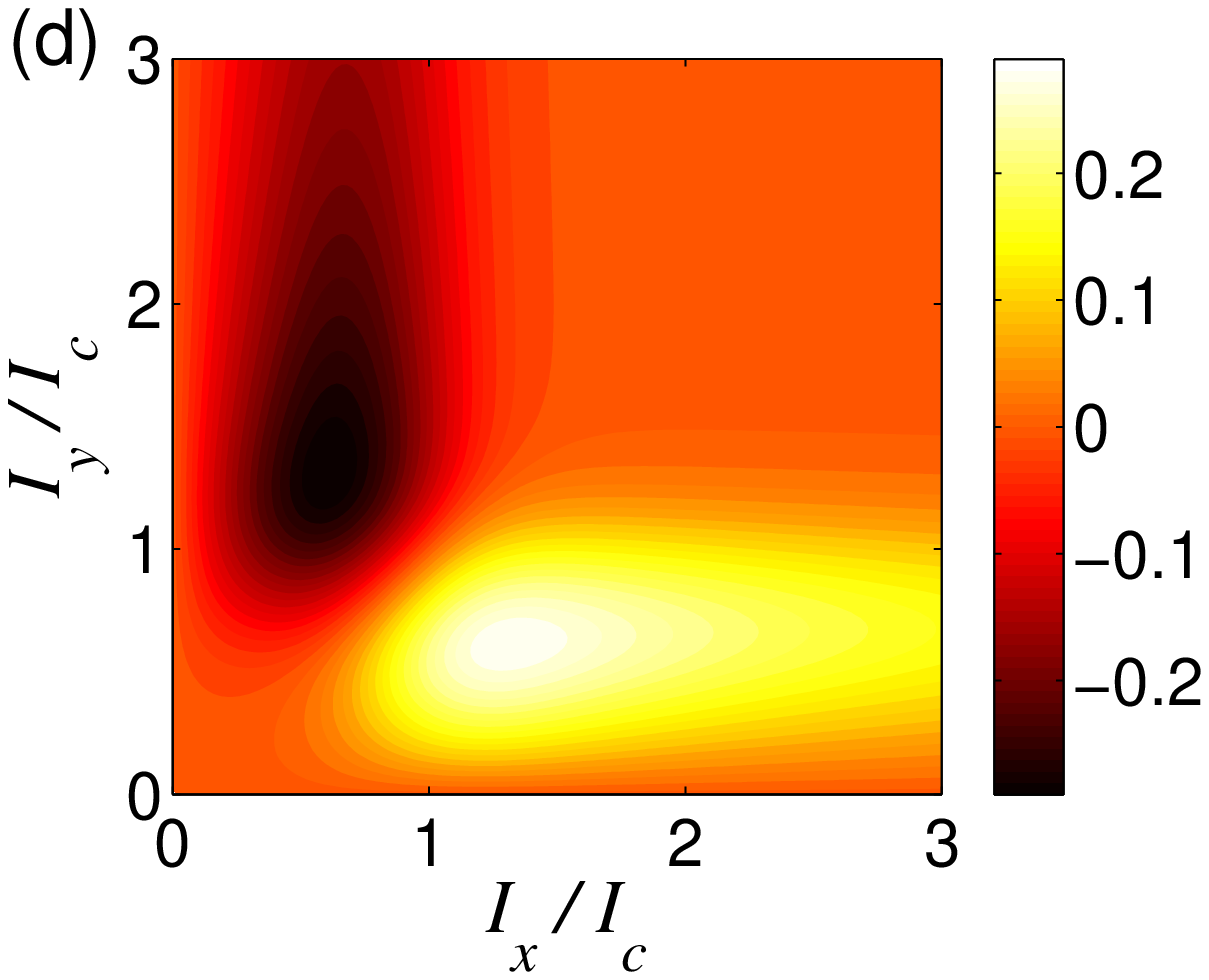}}
\end{tabular}
\caption{Longitudinal (a,c) and transverse (b,d) resistivity as a function of applied current $(J_x,J_y)$ for $p = 0.2$ (a,b) and  $p = 0.8$ (c,d).}
\label{fig:resist} 
\end{figure}

When current is applied in an arbitrary direction relative to the crystal axes, in addition to the longitudinal resistivity, there is also a finite transverse resistivity. 
The full angular dependence of the longitudinal and transverse resistivity obtained using Eq. (\ref{eq:EjTheta}) for $I_c  = 2$ and for $p = 0.2, \ 0.8$ is presented in Fig. \ref{fig:resist}.  As expected, the biggest anisotropy effects are concentrated in the range of currents where the resistivity of the superconducting links is nonlinear. 
This transverse response would be completely absent in a linear-resistivity of a tetragonal system. Therefore, it represents a   ``null-point measurement'' that can be a very sensitive diagnostic of the presence of local rotational symmetry breaking induced by superconducting stripes.

In order to relate the above model to superconducting cuprates we need to know how the parameters $(p, R_s^0, R_n, I_c)$ depend on the carrier concentration $x$ and temperature $T$.   Qualitatively, we expect that with the temperature increasing above $T_c$, the parameters $I_c$ and $p$ decrease, both vanishing at some temperature $T_c^*$, while $R_s^0$ and $R_n$ increase.   Experimentally, $T_c^*$  may correspond to the temperature at which the measured {\em linear} in-plane resistivity experiences a change of slope towards smaller values at lower temperatures, the $T_2^*$ of Ref.~\cite{Loram}. If this identification is correct, then below $T_2^*$, a non-linear resistivity behavior of the kind that we propose should exist.
We also expect that in the underdoped regime, at $T_c$, the system is near the percolation threshold, $p \approx 0.5$, with $p > 0.5$ at $T < T_c$ and $p< 0.5$ at $T > T_c$. We also note that the predicted non-linear transport signatures, are distinctly different from  another potential source of nonlinearity -- the Joule heating.  Namely, the window of currents where non-linear effects due to local superconductivity should exist, shrinks as the temperature increases above $T_c$, vanishing at some $T_c^*$, while Joule heating should increase with increasing current indefinitely and should be sensitive to the details of the thermal coupling to environment. Moreover, Joule heating is not expected to cause any spatial anisotropy in resistivity.

\begin{figure}[ht]
\begin{tabular}{ll}
{\includegraphics[width=1.7in]{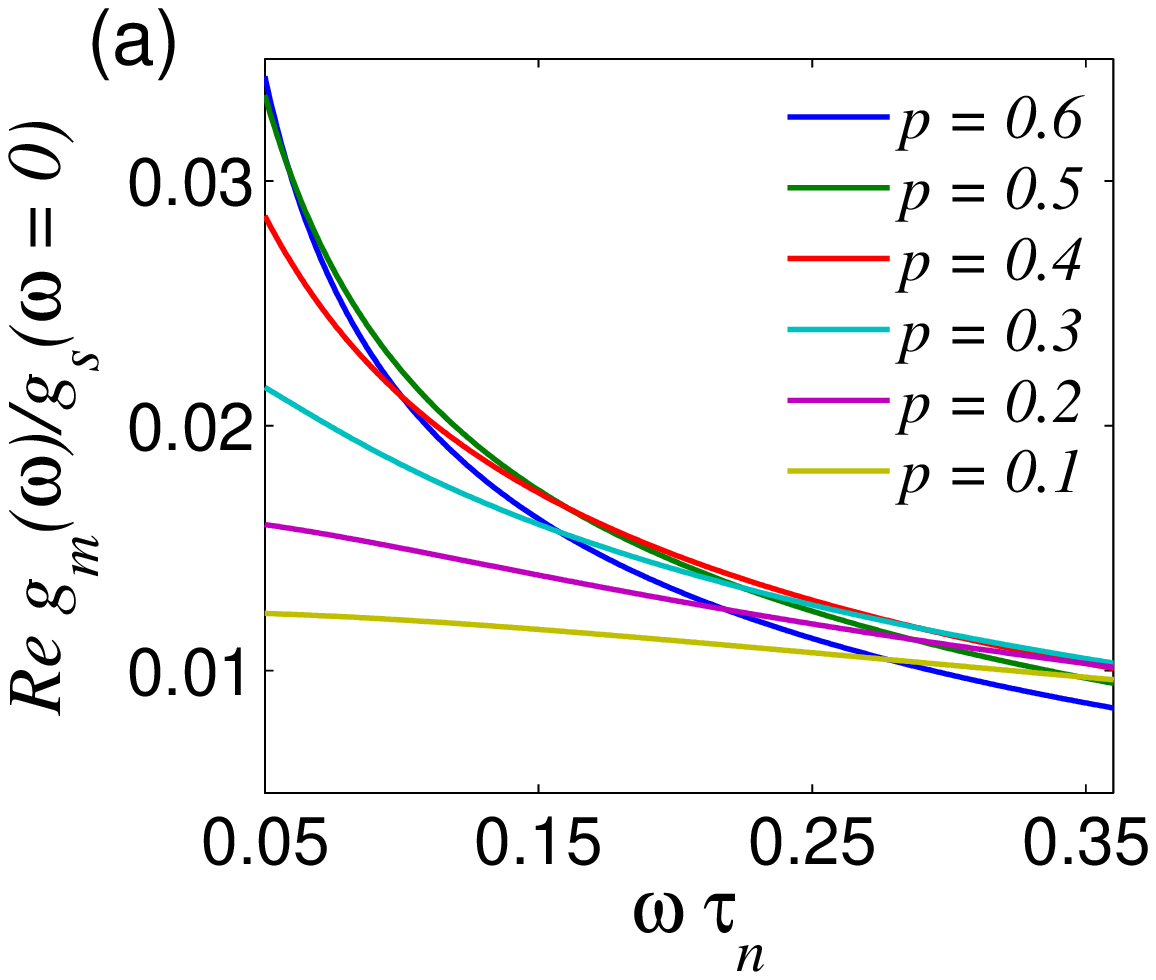}}  & {\includegraphics[width=1.7in]{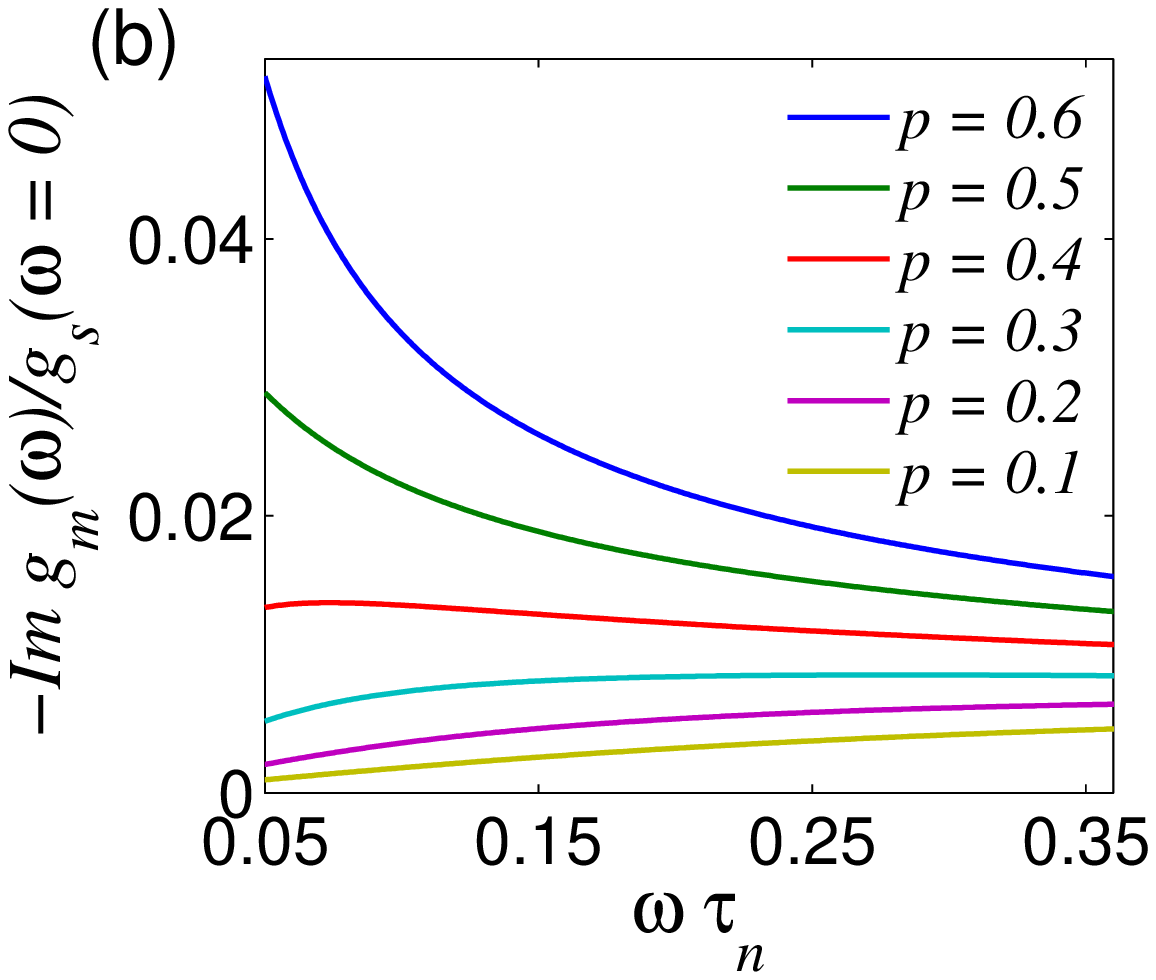}} \\
\end{tabular}
\caption{Frequency dependence of (a) Real and (b) Imaginary parts of conductivity obtained from effective medium theory for systems with different stripe concentrations $p$. $\tau_n/\tau_s = 100$.}
\label{fig:frequency} 
\end{figure}

Another issue that we can comment on within our superconducting percolation model is the apparent discrepancy between the superconducting fluctuation regime, as extracted from the bulk probes \cite{Ong} and from the AC transport measurements \cite{Oren2, Corson, Armitage}.  In the AC transport, one attempts to detect features characteristic of superconductivity, such as the superfluid density, and tracks their disappearance as a function of increasing temperature.
Experiments consistently find that superconducting fluctuation range above $T_c$, extracted this way, is much more narrow than the one obtained from the bulk measurements, such as diamagnetic response \cite{Ong}.  
Within the uniform fluctuating superconductor scenario these differences are difficult to reconcile \cite{Armitage}. 
However, if we assume that the superconductor is intrinsically inhomogeneous, as in the model considered above, the rapid disappearance of the {\em transport} signatures of superconductivity can be interpreted in terms of reduction of the superconducting fraction $p$ below the percolation threshold. 
Indeed, let us assume that the conductivity of the normal links is $g_n = \alpha \tau_n$ and of the superconducting links is $g_s = \alpha \tau_s/(1 + i\omega\tau_s)$ in the relevant frequency range, $\omega \tau_n\ll 1$, where $\tau_s$ is the relaxation time in the superconducting links and $\tau_n$ is the normal state relaxation rate ($\tau_s \gg \tau_n$). The effective medium conductivity $g_m$ can be obtained  from EMT, Eq. (\ref{eq:EMT}).  We find, above percolation threshold, $g_m \approx (2p -1) g_s$, while below -- $g_m \approx g_n/(1-2p)$, the latter having only a very small imaginary part. The crossover occurs in a narrow range of $\delta p \sim \sqrt{ \omega\tau_n}\ll 1$, i.e., the superconducting contribution disappears very rapidly as the fraction of the superconducting links goes below the percolation threshold $p = 1/2$ near $T_c$, even though {\em disconnected} superconducting inclusions may still persist to much higher temperatures and contribute to other superconductivity-sensitive measures.
In Fig. \ref{fig:frequency} we show schematically the behavior of $g_m(\omega)$ expected from the EMT, which agrees with the trends observed experimentally. 

The lattice percolation model presented here, while crude, provides a transport-based method to test the existence of local superconducting inclusions -- superconducting stripes -- in the cuprates at and above $T_c$.  In combination with the bulk probes, which are sensitive to the local  superfluid density ($\propto I_c$) and the volume fraction $p$, it may help to shed light on the nature of the pseudogap regime, from which the high-temperature superconductivity emerges.

We acknowledge useful discussions with P. Armitage, I. Bozovic, H. Lee, C. Panagopoulos, and T. Park.  This work was carried out under the auspices of the
National Nuclear Security Administration of the U.S. Department of Energy at Los Alamos National Laboratory under Contract No. DE-AC52-06NA25396 and supported
by the LANL/LDRD Program.

\end{document}